\documentclass[aps,12pt]{revtex4}
\usepackage{graphicx}
\usepackage{dcolumn}
\usepackage{bm}

\begin{document}

\title{Cold Atom Space Clock with Counter-Propagating Atoms}
\author{Desheng L\"u$^{1,2}$}
\author{Bin Wang$^{1,2}$}
\author{Tang Li$^1$}
\author{Liang Liu$^{1,*}$}

\affiliation{$^1$Key Laboratory of Quantum Optics, Shanghai
Institute of Optics and Fine Mechanics, Chinese Academy of
Sciences, Shanghai 201800, China}

\affiliation{$^2$Graduate School of the Chinese Academy of
Sciences, Beijing 100039, China}

\affiliation{$^*$Corresponding author: liang.liu@siom.ac.cn}

\date{\today}

\begin{abstract}
We discuss the feasibility to realize a space cold atom clock with
counter-propagating cold atoms in microgravity. The design of the
space clock is based on atomic beam clock with a Ramsey cavity,
except a magneto-optical trap (MOT) is placed at each side.  Cold
atoms are launched from MOTs at both side of the clock
simultaneously and move at counter-direction towards each other.
The velocity of launched atoms is precisely controlled to the
Ramsauer-Townsend resonance so that no additional collision
frequency shift is taken place. Such a configuration can
efficiently cancel the
frequency shift led from cavity phase shift and increase the signal to noise.\\
{\it OCIS codes:} 020.3320, 020.2070,

\end{abstract}
\maketitle
\noindent

In 1955, the first atomic frequency standard, using a Cs beam
excited by means of the separated oscillatory field, was completed
by Essen and Parry \cite{Essen55}. The approach was proposed by
Ramsey in 1950 \cite{Ramsey50}. In the Ramsey atomic clock, an
atomic beam is formed in an oven and are allowed to drift freely
in high vacuum into an interaction region formed by a microwave
structure called the Ramsey cavity.  The structure is generally
with a U-typed waveguide. This arrangement creates two short
microwave interaction regions of length $\ell$, separated by a
relatively large distance $L$.  After traversing the first
interaction region, atoms are exposed to the microwave field for a
short time, which depends on the distance $L$, and then enter the
second interaction region. Similar to the technique of magnetic
resonance, transitions of atoms are excited between the two
special levels. The advantage of the Ramsey type atomic clock is
that interferences take place between the excitation in the two
interaction regions, leading to a series of fringes which is
called ``Ramsey fringes". A narrow resonance can be obtained by a
factor of the order of $L/\ell$.

In experiment, it is very hard to keep the phases of microwave
between two interaction regions in an accurate consistent. A small
unwanted phase shift may be caused by an asymmetry in the Ramsey
cavity construction, and then the central fringe will be distorted.
The shift is so-called ``cavity phase shift". It is necessary to
cancel the cavity phase shift in theory or minimize it for improving
the accuracy of Ramsey type atomic clock.

In a classical thermal cesium clock, a beam of atoms effuses from
an oven and passes through a state-selecting magnet, and
subsequently passes through a Ramsey microwave cavity, and then
detected \cite{Bauch05}.  The velocity of thermal atoms can be as
slow as 95 m/s and line width of the clock transition is typically
around 60 Hz, as PTB Cs frequency standards apply \cite{Bauch05}.
In order to test the cavity phase frequency shift, in PTB's
primary clock CS2, an oven and detector is placed at each end so
that alternate operation of atomic beam in opposite directions can
be performed \cite{Bauch03-Sci}.

In 1954 Zacharias attempted to obtain an even narrower separated
oscillatory field resonance in a ``fountain" experiment
\cite{Ramsey05,Wynands05}.  The most important improvement in the
fountain clock is using one interaction field instead of two
regions. Atoms interact with the one interaction region two times
in the track of up and down, and interferences taking place
between twice interactions. Thus the cavity phase shift can be
cancelled in theory.  Unfortunately the experiment failed due to
the very slow atoms being scattered away as they emerged from the
thermal source.

The advent of laser cooling techniques open the door to a new
approach for the fountain clock \cite{Wynands05}. Atoms are first
captured and cooled in a MOT, and then launched upwards by a
technique called moving molasses. The width of Ramsey fringe for a
fountain is determined by $\Delta\nu =0.25\sqrt{g/2H} $, where $g$
is the gravitational acceleration, and $H$ is the maximum height
of launched atoms. Typically, for a cold atom fountain clock, the
width of the central Ramsey fringe is 1 Hz, which corresponds to
$H=0.3$ m. Narrower width is possible but technically difficult.
For example, $0.1$ Hz width requires $H=30$ m, which is
impractical.

Soon after the success of the fountain clock, people noticed that
even narrower width can be realized in microgravity environment
\cite{Laurent06}. In microgravity, the atoms move at constant
velocity after they are launched from a MOT, that means the slower
velocity of launched atoms leads to the longer interrogation time,
or the narrower width of central Ramsey fringe.  In micro-gravity
environment, however, the design of the fountain clock can not be
adopted in the cold atom space clock, and Ramsey cavity structure
is recalled again, and thus cavity phase shift should be focused
again especially for a cold atom space clock with high accuracy
and stability.

In cold atom space clock, the PHARAO \cite{Laurent06} for example,
cold atoms are launched from a MOT at velocity as low as 5 cm/s,
and $0.1$ Hz width of central Ramsey fringes is predicted. The
expected stability is $10^{-13}/\sqrt{\tau}$, where $\tau$ is the
integration time, and the accuracy is up to $10^{-16}$. For such a
high stability and accuracy, the phase shift of the space clock's
Ramsey cavity becomes more important. Certainly, it is possible to
apply PTB's CS2 clock design in the space clock, and alternate
operation of cold atoms in opposite direction gives the
information of the cavity phase shift, but such a design wastes
precious space resources.

\begin{figure}
\centerline{\includegraphics[width=3.3in,height=1.3in]{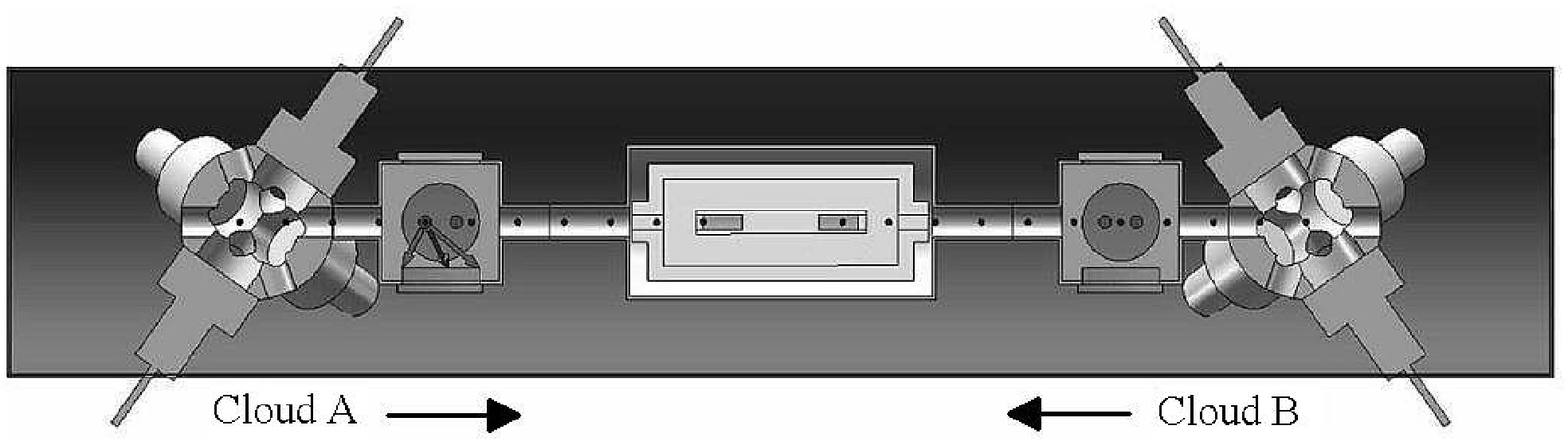}
} \caption{Schematic diagram of cold atom space clock with
counter-propagating atoms} \label{fig:1}
\end{figure}

In this paper, we propose a new type of space clock, whose design
is similar to the PTB's CS2 \cite{Bauch03-Sci}, but with
completely new operation mode. The new type of space clock aims at
cancelling the frequency shift due to the phase difference of the
Ramsey cavity and increasing the signal to noise ratio, and thus
reducing the technical difficulties of the design and improving
the performance of the space clock. As shown in Fig. \ref{fig:1},
a MOT and a detection region is placed at each end. Cold atoms are
launched from both MOTs simultaneously in opposite direction
towards each other. Assuming Cloud A denotes the cold atoms
launched from the left MOT, and Cloud B from the right MOT, Cloud
A collides with Cloud B at the center of the Ramsey cavity after
they pass through the first interaction region of the cavity.

In atom-atom collisions, if the atoms are treated classically as
hard balls, the calculated cross section is independent of the
atomic energy. In quantum mechanics, however, the atoms are
considered to present a dipole-dipole interaction of typical
atomic dimensions for the scattering between atoms. The solution
of Schr\"odinger Equation for two dipole-dipole potentials shows
that the cross-section in atom-atom collisions will have a minimum
at some specific energies. This is a simple illustration of the
Ramsauer-Townsend resonance \cite{Mott-Massey}. At the
Ramsauer-Townsend resonance, the atoms are transparent with each
other when they collide, and frequency shift lead from collisions
becomes null. Gibble {\it et.al.} measured the $s$-wave frequency
shift with juggling $^{133}$Cs and $^{87}$Rb fountain clock
\cite{Legere98,Fertig00,Fertig01}. The first Ramsauer-Townsend
shift null for $^{87}$Rb happens at alternate launch time delays
of $\Delta t=22$ ms between two cold atom balls, corresponding to
the velocity $v_{RT}=g\Delta t/2=10.78$ cm/s of each ball
propagating at opposite direction. At such a velocity, the
collision between the Cloud A and B does not contribute additional
frequency shift. For the scattering of identical particles, the
$p$-wave scattering does not contribute additional frequency
shift.

We assume that the interrogation length of the Ramsey cavity is
$L=52.5$ cm, which gives the interrogation time
$T_{RT}=L/v_{RT}=4.9$ s when Cloud A or Cloud B moves at the
Ramsauer-Townsend velocity $v_{RT}$. Such an interrogation time
corresponds to the linewidth of central Ramsey fringe at $\Delta
\nu =0.1$ Hz. If the number of detected atoms is $N=10^6$, we have
the Allan variance
\begin{equation}\label{allan}
\sigma_y(\tau)={\Delta \nu \over \pi \nu_0 \sqrt{N}}\sqrt{{T\over
\tau}}=1.5\times 10^{-14}/\sqrt{\tau}
\end{equation}
where, $T=10$ s which includes time for the preparation of cold
atoms, state selection, interrogation time and detection, and
$\nu_0 =6.835$ GHz is the frequency of clock transition of
$^{87}$Rb, and $\tau$ is the integration time. If we do not
consider the phase shift of the Ramsey cavity, we can average the
detected atoms from both Cloud A and B such that $N=2\times 10^6$,
which leads to the Allan variance to be reduced by a factor of
$\sqrt{2}$ from Eq. (\ref{allan}), that is
$\sigma_y(\tau)=1.1\times 10^{-14}/\sqrt\tau$.

On the other hand, with atoms' counter-propagating through a
Ramsey cavity, the cavity phase shift can be accurately measured,
and thus the frequency shift can be adjusted away from the error
budgets of the clock. Or the frequency shift due to phase
differences of the cavity can be cancelled if we average the
signals from both Cloud A and Cloud B. Assuming a phase difference
$\Delta\varphi$ between two zones of the Ramsey cavity, we can
easily have the probability to find the two-level system in the
excited state as \cite{Riehle_book}
\begin{eqnarray}
p_A={1\over 2}\sin^2\Omega t\{1+\cos[2\pi(\nu -\nu_0) T_{RT}+\Delta\varphi]\}\label{p_a}\\
p_B={1\over 2}\sin^2\Omega t\{1+\cos[2\pi(\nu -\nu_0)
T_{RT}-\Delta\varphi]\}\label{p_b}
\end{eqnarray}
Here, $p_A$ and $p_B$ are the probability for Cloud A and Cloud B,
respectively. $\Omega$ is the Rabi frequency, $t$ is the
interaction time between atoms and microwave. Generally, the phase
difference $\Delta\varphi$ shifts the center of Ramsey fringe by
\begin{equation}\label{delta_nu_phi}
{\Delta\nu_\varphi \over \nu_0} =-{\Delta \varphi \over 2\pi \nu_0
T_{RT}}
\end{equation}
in $p_A$ and $-\Delta\nu_\varphi /\nu_0$ in $p_B$. Typically, the
phase difference of a U-type Ramsey cavity can be controlled below
a few hundred $\mu$rad. If we take the phase difference $\Delta
\varphi=500$ $\mu$rad for example, we have ${\Delta\nu_\varphi /
\nu_0}=2.3\times 10^{-15}$. Thus in order to get accuracy of a few
$10^{-16}$, the frequency shift due to the phase difference of the
cavity must be carefully considered.

If we take an average over the probability of Cloud A and Cloud B,
we have
\begin{eqnarray}
p&=&{p_A+p_B\over 2} \nonumber \\
 &=&{1\over 2}\sin^2\Omega
t[1+\cos2\pi(\nu-\nu_0) T_{RT}\cdot\cos \Delta\varphi]
\end{eqnarray}
Obviously, in $p$, the phase difference of the cavity does not
contribute any frequency shift, but the width of the central
Ramsey fringe has been broadened to
\begin{equation}\label{delta_nu_p}
\Delta\nu_p=\Delta\nu+2\Delta \nu_\varphi
\end{equation}
Typically phase difference of the Ramsey cavity is around a few
hundred $\mu$rad \cite{Bauch05}. From Eq. (\ref{delta_nu_phi}), we
have $\Delta \nu_\varphi\approx 1.6\times 10^{-5}$ Hz when $\Delta
\varphi=500$ $\mu$rad for example, which can be neglected in Eq.
(\ref{delta_nu_p}), and thus the width broadening due to the
average of the signal from both counter-propagating atoms can be
neglected in the Allan variance given in Eq. (\ref{allan}). Since
the collision between Cloud A and Cloud B, each moves at the
Ramsauer-Townsend velocity, does not contribute additional
frequency shift, the frequency shift due to the cavity phase
difference in our system can be cancelled without additional cost.

In addition, our design has a very important function for reducing
the noise of the interrogation time due to the vibrations on the
space craft and variations of residual microgravity in the atoms'
propagating direction. This feature was first realized by Fertig
{\it et al.} in the design of a microgravity atomic clock with the
double Ramsey cavity \cite{Fertig00-RACE}. Fertig's design has two
Ramsey cavities located at both sides of a MOT, and cold atoms are
launched alternatively to each Ramsey cavity. Since the
alternatively launching of cold atoms in the opposite direction
does not happen at the same time, the cold atoms do not sense the
same vibration and residual microgravity. In our design, however,
the counter-propagating atoms through same cavity at the same
time, the detected signals of Cloud A and Cloud B behave
oppositely and simultaneously and thus the effect of vibrations
and variations of the microgravity can be removed by the averaged
signal $p$.

Our design has some unique features especially suitable for some
experiments. For example, if the velocity of launched atoms
varies, the cross-section of atom-atom scattering can be measured
by our space clock even more precisely than by the juggling
fountain clock. Such a measurement can give detailed data for
atom-atom scattering and test the fundamental principle of quantum
mechanics.

In conclusions, we have proposed a new type of cold atom space
clock with counter-propagating atoms. The cold atoms move at
Ramsauer-Townsend velocity so that the collision between
counter-propagating atoms becomes null. We pointed out that such a
null collision cross section can cancel the frequency shift lead
from the phase difference of the Ramsey cavity, and also increase
the signal to noise ratio. We estimated the Allan variance of such
a space clock at $1.5\times 10^{-14}/\sqrt\tau$. Besides, our
design of the space clock can efficiently remove the noise due to
the vibration and residual microgravity of the space craft, and
thus reduce the requirement of space environment.

This work is supported by the National High-Tech Programme under
Grant No. 2006AA12Z311 and National Basic Research Programme of
China under Grant No. 2005CB724506. The authors would like to
thank Professor Shougang Zhang and Professor Wuming Liu for their
valuable discussions.

\newpage

\end{document}